\documentstyle[12pt]{article}
\def\nn{\noindent}
\def\nnb{\nonumber}
\def\lf{\left(}
\def\rg{\right)}

\def\bs{\bar{s}}
\def\smn{\sigma^{\mu\nu}}
\def\cO{\cal{O}}
\def\cT{{\cal{T}}}

\def\bbbone{{\mathchoice {\rm 1\mskip-4mu l} {\rm 1\mskip-4mu l}
{\rm 1\mskip-4.5mu l} {\rm 1\mskip-5mu l}}}

\def\beq{\begin{equation}}
\def\eeq{\end{equation}}
\def\bea{\begin{eqnarray}}
\def\eea{\end{eqnarray}}

\def\bbbone{{\mathchoice {\rm 1\mskip-4mu l} {\rm 1\mskip-4mu l}
{\rm 1\mskip-4.5mu l} {\rm 1\mskip-5mu l}}}
\def\bbbz{{\mathchoice {\hbox{$\sf\textstyle Z\kern-0.4em Z$}}
{\hbox{$\sf\textstyle Z\kern-0.4em Z$}}
{\hbox{$\sf\scriptstyle Z\kern-0.3em Z$}}
{\hbox{$\sf\scriptscriptstyle Z\kern-0.2em Z$}}}}
\def\npb#1#2#3{    {\it Nucl. Phys. }{\underbar{B#1}} (19#2) #3}
\def\plb#1#2#3{    {\it Phys. Lett. }{\underbar{B#1}} (19#2) #3}

\def\ib#1#2#3{     {\it ibid. }{\underbar{#1}} (19#2) #3}
\relax
\begin{document}
\centerline{ {\bf $b \to s \gamma$ \ AND \ MINIMAL
                                   \ SUPERSYMMETRY       }
                                                   \footnote{
To be published in the Proceedings of {\it XV Meeting on Elementary
   Particle Physics}, Kazimierz, Poland, May 1992.}     }

\vspace*{1.0cm}
\centerline{\bf Francesca M. Borzumati}
\vspace*{0.2cm}
\centerline{{\it II.\ Institut f\"ur Theoretische Physik
\footnote{Supported~by~the~Bundesministerium~f\"ur~Forschung~und
           Technologie,~05~5HH~91P(8),~Bonn,~FRG.   } }}
\centerline{{\it Universit\"at Hamburg, 2000 Hamburg 50,
 Germany}}
\vspace*{1.0mm}
\begin{center}
\parbox{13.5cm}
{\begin{center} ABSTRACT \end{center}
\vspace*{-1mm}
{\small
\nn The contribution to the decay $b\to s \gamma$ in the Minimal
Supersymmetric Standard Model with radiative electroweak breaking
is presented. The impact of a detection of this inclusive decay at
the SM level is discussed. Results for other inclusive $b-s$
transitions are also mentioned. }}
\end{center}

The data collected in the last few years at CESR and DORIS~II have
sensibly reduced the upper bounds on some rare B processes. In
particular, the CLEO vertex detector allows to put a limit on the
inclusive decay $b \to s \gamma $, thus circumventing the still
sizeable theoretical uncertainty in the evaluation of the matrix
element $<K^*|\bs_{\rm L} i \smn b_{\rm R}|B>. $ The present limit on
$BR(b \to s \gamma)$ of $8.4 \times 10^{-4}$ \cite{DANIL}, only about
a factor of two away from the Standard Model (SM) prediction for
$m_t \simeq 130\,$GeV \cite{ODONN}, makes
therefore the detection of this process quite imminent.
 \\[1.2ex]
\indent
It comes then natural to ask oneself where the past claims of
sensitivity of rare B processes to new physics stand today. The first
point to be inquired is whether these claims have survived the steady
increase in the experimental lower bounds for the mass of particles
predicted by models beyond the standard one. In case they have, then,
the other question to be addressed is what type of restriction on the
allowed parameter-space of these models can be enforced by a detection
of $b \to s \gamma$ at the SM level.
The aim of this talk is to answer these two questions in the case of
Supersymmetry (SUSY). In this framework, a possible enhancement with
respect to the SM predictions had been claimed in the past for the
inclusive radiative decays $b\to s \gamma$ and $b \to s g$, while no
sizeable deviation from the SM level had been considered possible for
other flavour changing processes as $b \to s q \bar{q}$,
$b \to s \ell^+\ell^-$, $b \to s \nu\bar{\nu}$,
and for the oscillations $B_d^0 -\bar{B}_d^0$, $B_s^0 -\bar{B}_s^0$
\cite{USBSGAM}.
 \\[1.2ex]
\indent
The reasoning behind these claims can be quickly explained. Before
doing so, I shall list here the effective operators relevant for the
previous $b-s$ transitions and briefly recall the impact of QCD
corrections on the relative Wilson coefficients. The two-quarks
effective operators
contributing to the radiative decays $b \to s \gamma$, $b \to s g$ are:
\bea
{\cO}_{{\rm LR}}^{ph}        & \equiv &
 \lf \bs_{\rm L} i \smn b_{\rm R}\rg m_b \ {\rm q}_\nu \epsilon_\mu
                                                        \nnb \\
{\cO}_{{\rm LR}}^{gl}        & \equiv &
\lf \bs_{\rm L} i \smn T^a b_{\rm R}\rg m_b {\rm q}_\nu \epsilon_\mu^a
\label{operatone}
\eea
where $\sigma^{\mu\nu}\equiv (i/2)[\gamma^\mu,\gamma^\nu]$;
$\epsilon_\mu$ and $\epsilon_\mu^a$ are photon and gluon fields,
respectively; $T^a$ is a generator of $SU(3)$; ${\rm q}_\nu$ is the
momentum of the emitted gauge boson and finally the subindices
${\rm L}$,${\rm R}$ indicate the left- or right-handedness of each
fermion. Similarly, the four-quarks effective operators contributing
to the remaining set of processes mentioned before can be divided in
the groups:
\bea
{\cO}_{{\rm (LL)V\phantom{L)}}}   & \equiv &
     \lf \bs_{\rm L} \gamma^{\mu} \cT b_{\rm L}\rg
     \lf \bar{f} \gamma_{\mu} \cT f \rg                  \nnb  \\
{\cO}_{{\rm (LR)V\phantom{L)}}}    & \equiv &
     \lf \bs_{\rm L} i \smn \cT b_{\rm R}\rg
       m_b\frac{{\rm q}_\nu}{{\rm q}^2}
     \lf \bar{f} \gamma_{\mu} \cT f \rg                  \nnb  \\
{\cO}_{{\rm (LL)(LL)}}              & \equiv &
     \lf \bs_{\rm L} \gamma^\mu \cT b_{\rm L}\rg
     \lf \bar{f}_{\rm L} \gamma_{\mu} \cT f_{\rm L}\rg   \nnb  \\
{\cO}_{{\rm (LL)(RR)}}              & \equiv &
     \lf \bs_{\rm L} \gamma^\mu \cT b_{\rm L}\rg
     \lf \bar{f}_{\rm R} \gamma_{\mu} \cT f_{\rm R}\rg
\label{operattwo}
\eea
where $f$ is a generic fermion $q$, $\ell$, $\nu$ etc.; the matrix
$\cT$ (which can be the identity matrix $\bbbone$ or a generator $T^a$)
accounts for a possible colour structure of the two currents in each
operator, and the subindex ${\rm V}$ is used to denote a vector-current.
Subleading operators proportional to the $s$-quark mass are neglected
and all the coupling constants are also omitted, for simplicity.
Penguin diagrams with exchange of massless bosons, i.e. photons and
gluons (${\rm q}_\nu$ is the momentum of the exchanged boson),
contribute to the operators ${\cO}_{{\rm (LL)V}}$ and
${\cO}_{{\rm (LR)V}}$, while Z-mediated penguins contribute to
${\cO}_{{\rm (LL)V}}$ and ${\cO}_{{\rm (LL)(LL)}}$. Box diagrams
account for ${\cO}_{{\rm (LL)(LL)}}$ and ${\cO}_{{\rm (LL)(RR)}}$. The
last operator, as well as ${\cO}_{{\rm (LL)(LL)}}$ with $\cT=T^a$, not
present in the SM, arise in the supersymmetric case from box diagrams
with virtual Majorana fermions (gluinos $\widetilde{g}$) running in the
loop.
 \\[1.2ex]
\indent
The special role played by the operator ${\cO}_{{\rm LR}}^{ph,gl}$ and
${\cO}_{{\rm (LR)V}}$ within the SM, was quickly realized as soon as
interest started gathering around these flavour changing processes. It
was found that, due to mixings with the operators
$ \lf \bar{c}_{\rm L} \gamma^\mu b_{\rm L}\rg
  \lf \bar{s}_{\rm L} \gamma_\mu c_{\rm L}\rg $,
the Wilson coefficients relative to ${\cO}_{{\rm LR}}^{ph,gl}$ and
${\cO}_{{\rm (LR)V}}$ undergo a substantial strong correction during
the evolution from $M_W$ to $m_b$ (for a discussion of these issues
see \cite{ODONN} and references
therein). These coefficients, with a typical implementation of the
GIM mechanism of power-type, $\sim m_t^2/M_W^2$, roughly speaking,
acquire additive
logarithmic terms $\sim \ln(m_t^2/m_c^2)$. The effect is a
significant change of their value on one side, and a reduction of
their sensitivity to the top quark mass,
$m_t$, on the other. Indeed, an enhancement bigger than 100\% can be
found for the decay $b \to s \gamma$, while destructive interferences
lead to a suppression of $b \to s g$.
 \\[1.2ex]
\indent
The situation is more
complicated for $b \to s \ell^+\ell^-$. The strong enhancement relative
to ${\cO}_{{\rm (LR)V}}$ makes the contribution of this operator to
$\Gamma (b \to s \ell^+\ell^-)$ roughly
of the same size as the contributions
due to ${\cO}_{{\rm (LL)V}}$ and ${\cO}_{{\rm (LL)(LL)}}$. The QCD
correction to the photon-mediated penguin contributing to
${\cO}_{{\rm (LL)V}}$, already exhibiting a logarithmic GIM
suppression, do not have the same impact they had in the case of
${\cO}_{{\rm (LR)V}}$. Finally, Z-mediated penguins and box diagrams,
with a power-type implemention of the GIM mechanism, do not obtain
strong corrections. The net result is a modest enhancement
of the inclusive decay $b \to s \ell^+\ell^-$, never exceeding 20\%.
No correction is present in $b \to s \nu \bar{\nu}$ and related
decay modes as $B_s \to \tau^+ \tau^-$. Similar to
$b \to s \ell^+ \ell^-$ is the case of the transition
$b\to s q\bar{q}$. Although potentially interesting since it can
give together with $b \to s g g $ (and $b \to s g$) non-trivial
rates to hadronic decays of the $B$ meson, as $B\to K_S \phi$, I
shall neglect it in the following discussion. A re-analysis of
the supersymmetric
contribution to the non-charmed hadronic decay $b \to s q \bar{q}$,
in fact, has not been performed; it exists  only for $b \to s g$.
Finally, suppression factors around 15\% are obtained for the
mixings $B^0-\bar{B}^0$.
 \\[1.2ex]
\indent
The reason for dwelling so long upon the type of modifications
that QCD corrections bring to each operator, is that one can
draw interesting analogies with the corrections due to the exchange
of supersymmetric particles in the internal loops. Again,
${\cO}_{{\rm LR}}^{ph,gl}$ and ${\cO}_{{\rm (LR)V}}$ are, in
principle, the likely operators to induce a sizeable enhancement of
supersymmetric signals over the SM ones. The explanation is quite
simple. For long time, the effective tree-level coupling
$\widetilde{g}-q-\widetilde{q}^{\ \prime}$, with quarks and squarks,
$q$, $\widetilde{q}^{\ \prime}$, of different flavours, was considered
the key ingredient to induce flavour change in quark transitions. In
this case, in fact, the presence of the strong coupling $\alpha_{\rm S}$
versus the
weak one, typical of the SM, produces a net gain for SUSY if all the
other ingredients entering in the loop calculation, contribute
numerically in the same way in the two frameworks. Even assuming
that the generational splitting among squarks is of the same size
as the one among the quarks running in the SM loop, the higher
average squark mass gives a stronger squark degeneracy and,
numerically, a supersymmetric GIM suppression {\it at least} as
effective as the power-type suppression. Therefore, the correct
strategy is to consider only processes involving operators which
already have a strong GIM suppression in the SM. Since Z-mediated
penguins and box diagrams can be excluded as unlikely
possibilities, see \cite{USBIG}, the only viable operators one is left
with are ${\cO}_{{\rm LR}}^{ph,gl}$ and ${\cO}_{{\rm (LR)V}}$.
The expectations are then shaped on the results obtained in the
case of the strong corrections: while a possible enhancement of
the supersymmetric contribution of ${\cO}_{{\rm (LR)V}}$ to
$b \to s \ell^+\ell-$ would be diluted by the contributions
coming from different operators, the radiative processes
$b\to s \gamma$, $b \to s g$ can give, in principle, a clear
signature of SUSY above the SM. These are the conclusions reached
in \cite{USBSGAM} for squarks and gluinos masses taken at the weak
scale and when the abovementined splitting among squarks of
different generations is assumed.
 \\[1.2ex]
\indent
However, it is clear that the increase of the experimental
lower bounds of supersymmetric particles does not allow to
single out the effective vertex
$\widetilde{g}-d-\widetilde{d}^{\ \prime}$ as the dominant source of
flavour change and that all the possible
sources contributing to $b-s$ transitions have to be considered. They
are distinguishable according to the virtual
particles running in the loop mediating the $b-s$ transition:
1) $W^-$boson and $u$-quarks;
2) charged Higgs $H^-$ and $u$-quarks;
3) charginos $\widetilde\chi^-$ and $u$-squarks;
4) neutralinos $\widetilde\chi^0$ and $d$-squarks;
5) gluinos $\widetilde g$ and $d$-squarks.
 \\[1.2ex]
\indent
The calculation of all these contributions and their possible
interference effects can be reasonably performed within the minimal
supersymmetric standard model with radiative breaking of the
electroweak sector. Apart from $m_t$ and $\tan \beta$ (the ratio of
vacuum expectation values giving rise to the $u$- and $d$-quark
masses), only two new parameters are needed to fully specify this
model. We choose them to be $m$ and $M$, the common masses which
all scalars and gauge fermions acquire respectively at the Planck
scale, after the soft breaking of supersymmetry. Once the value of
these parameters is fixed, the correct boundary conditions at a grand
unified scale are kept in account, and the low-energy input values
for $\alpha$, $\alpha_{\rm S}$, $\sin^2 \theta_W$ and $m_b$ are
considered, all mass parameters and couplings present in the model
can be calculated. They are obtained by integrating the relative
renormalization group equations and by requiring that
the scalar potential acquires at the electroweak scale the minimum
needed for the correct breaking of
$SU(2)_{\rm L} \times U(1)_{\rm Y}$. The details of the procedure
followed for this analysis can be found in \cite{USBIG,MESUSY}.
 \\[1.2ex]
\indent
It turns out that the constraint of radiative breaking of
$SU(2)_{\rm L} \times U(1)_{\rm Y}$ is rather stringent.
Interesting relations among the supersymmetric masses are obtained,
depending on the particular values of $m_t$, $\tan \beta$, $m$ and
$M$. One feature, though, is quite general and has strong
consequences on the size of the various contributions to the
$b-s$ transitions discussed here: the lightest $d$-squark,
$\widetilde{d}_1$, is in general heavier than the lightest
$u$-squark, $\widetilde{u}_1$. Moreover, the implementation of a
modest lower bound on $m_{\widetilde{u}_1}$ ($25-30\,$GeV) can push
$m_{\widetilde{d}_1}$ towards much higher values. This fact, together
with the requirement that $m_{\widetilde{g}}$ is above $100-150\,$GeV,
strongly suppresses the size of the gluino contribution to
$b \to s \gamma$. The neutralino contribution, further penalized by
the smallness of its coupling, is then completely negligible. The two
biggest supersymmetric contributions come from the exchange of $H^-$
and $u$-quarks and of $\widetilde{\chi}^-$ and $u$-squarks. The
elements playing an important role are obviously the presence of the
top-Yukawa-coupling in the first case and the contribution of the
lightest squark $\widetilde{u}_1$ in the second one. However, both
contributions are below the SM prediction: at most $60\%$ of the
contribution coming from the $W^-$ and $u$-quarks exchange
can be obtained by the exchange of $H^-$ and $u$-squarks for
$m_t=130\,$GeV and $\tan \beta=2$.
%
These contributions add constructively to give a band of
supersymmetric results almost completely above the SM prediction, as
shown in Fig.~1 for $\tan \beta = 2,8$ and $m_t=130\,GeV$. The width
of this band is due to the dependence on the remaining supersymmetric
parameter $M$. QCD corrections are also implemented in these results
(for details see \cite{USBIG}). An enhancement of a factor $2-3$,
is obtained in the case of $b \to s \gamma$.

As expected, $b \to s \nu \bar{\nu}$ and $B_s \to \tau \bar{\tau}$ do not
show any deviation from the  SM prediction and the same is true for the
oscillations $B^0-\bar{B}^0$ \cite{USBIG}. Interesting is instead the
result obtained for $b \to s \ell^+\ell^-$. The shape of the
supersymmetric band of values for $BR(b \to s \ell^+\ell^-)$, similar to
the one for $BR((b \to s \gamma)$, is a  clear indication that the
enhancement observed in this case is due to the operator
${\cO}_{{\rm (LR)V}}$. This enhancement, though, would not be
visible if the QCD corrections to this operators had not increased its
contribution to $BR(b\to s \ell^+\ell^-)$ and if destructive
interferences among the various components contributing to the
remaining operators had not occurred.
Although the enhancement factor over the SM prediction is similar to
the one obtained for $b\to s \gamma$, this decay is much less
interesting at the moment since it is still quite far from experimental
detection.

\vspace{0.5cm}
\vspace*{4.0cm}
\begin{small}
\nn \begin{minipage}[c]{16cm}{{\bf Fig.~1} Branching Ratios of various
loop induced $b-s$ transitions in the Minimal Supersymmetric
Standard Model for $m_t=130\,$GeV and $\tan \beta =2,8$, as a
function of the soft breaking scalar mass $m$. The horizontal lines
correspond to the SM predictions ($f_{B_s}=150\,$MeV). From
ref.~\cite{USBBOOK}.   }
\end{minipage}
\end{small}
\vspace*{0.5cm}

As for $b \to s \gamma$, the enhancement obtained in the framework
of Minimal Supersymmetry with radiatively induced breaking of
$ SU(2)_{\rm L}\times U(1)_{\rm Y}$ is smaller than the values
claimed in the past.
However, even this reduced enhancement makes $b\to s \gamma$
sensitive to region in the supersymmetric parameter space not yet
excluded by collider searches.
As can be seen in Fig.~2, the experimental detection of
$BR(b\to s \gamma)$ at 1.5 times the SM prediction for
$m_t=130\,$GeV and
$\tan\beta$ can induce restrictions of the
parameter space  competitive with the ones due to
negative results in searches of supersymmetric particles at
LEP\,I and LEP\,II.

\vspace{0.5cm}
\vspace*{4.0cm}
\begin{small}
\nn \begin{minipage}[c]{16cm}{{\bf Fig.~2} Comparison between
the regions of the plane $m_{\widetilde{g}}-\mu$ excluded at
LEP\,I (area enclosed by the thick solid line), the projected
limits obtainable at LEP\,II (area below the dashed line) and
the regions excluded by: i) the requirement of radiative
electroweak breaking (A); ii) a bound on
$BR(b \to s \gamma) \ 50\%$ above the SM prediction (A+B);
iii) $15\%$ above the SM prediction (A+B+C).
{}From ref.~\cite{USBBOOK}  }
\end{minipage}
\end{small}
\vspace*{0.5cm}
\begin{small}

\end{small}

\end{document}